\documentclass[prd,nofootinbib,showpacs,12pt]{revtex4}
\usepackage{amsmath}

\begin{document}

\title{Spherically symmetric problem on the brane \\ and galactic rotation
curves}

\author{Alexander Viznyuk}\email{viznyuk@bitp.kiev.ua}
\author{Yuri Shtanov}\email{shtanov@bitp.kiev.ua}
\affiliation{Bogolyubov Institute for Theoretical Physics, Kiev 03680, Ukraine}


\begin{abstract}
We investigate the braneworld model with induced gravity to clarify the role of
the cross-over length scale $\ell$ in the possible explanation of the
dark-matter phenomenon in astrophysics and in cosmology. Observations of the
21~cm line from neutral hydrogen clouds in spiral galaxies reveal that the
rotational velocities remain nearly constant at a value $\upsilon_c\sim
10^{-3}$$-$$10^{-4}$ in the units of the speed of light in the region of the
galactic halo. Using the smallness of $\upsilon_c$, we develop a perturbative
scheme for reconstructing the metric in a galactic halo.  In the leading order
of expansion in $\upsilon_c$, at the distances $r\gtrsim \upsilon_c\ell$, our
result reproduces that obtained in the Randall--Sundrum braneworld model. This
inequality is satisfied in a real spiral galaxy such as our Milky Way for
distances $r \sim 3$~kpc, at which the rotational velocity curve becomes flat,
$\upsilon_c \sim 7\times 10^{-4}$, if $\ell \lesssim 2$~Mpc. The gravitational
situation in this case can be approximately described by the Einstein equations
with the so-called Weyl fluid playing the role of dark matter. In the region
near the gravitating body, we derive a closed system of equations for static
spherically symmetric situation under the approximation of zero anisotropic
stress of the Weyl fluid. We find the Schwarzschild metric to be an approximate
vacuum solution of these equations at distances $r\lesssim
\sqrt[3]{r_g\ell^2}$. The value $\ell \lesssim 2$~Mpc complies well with the
solar-system tests. At the same time, in cosmology, a low-density braneworld
with $\ell$ of this order of magnitude can mimic the expansion properties of
the high-density LCDM ($\Lambda$ + cold dark matter) universe at late times.
Combined observations of galactic rotation curves and gravitational lensing can
possibly discriminate between the higher-dimensional effects and dark matter.
\end{abstract}



\maketitle

\vfill\eject

\section{Introduction}

The idea that our familiar four-dimensional space-time is a hypersurface
(brane) of a five-dimensional space-time (bulk) \cite{ADD,RS,DGP} was under
detailed elaboration during the last decade. According to this braneworld
scenario, all matter and gauge interactions reside on the brane, while gravity
can propagate in the whole five-dimensional space-time. An observer in this
scenario is in direct contact only with the induced metric on the brane.

A ponderable argument in favor of the braneworld theory is the ability to solve
outstanding problems of modern cosmology and astrophysics. The generally
accepted LCDM ($\Lambda$ + cold dark matter) model is in a good agreement with
most of experimental data; however, up to now, no non-gravitational evidence
for dark matter has been reliably found.  It was argued \cite{MH,PBK,HC,BH}
(see also \cite{GogMaz} for a somewhat different approach) that a modified
theory of gravity based on the Randall--Sundrum braneworld scenario \cite{RS}
can explain the observations of the galactic rotation curves, while
observations of the gravitational lensing can possibly discriminate between the
higher-dimensional effects and dark matter. In a recent work \cite{HC7}, the
authors argued that the virial-theorem mass discrepancy in clusters of galaxies
can also be accounted for in the Randall--Sundrum model. However, this version
of braneworld model cannot explain the dark-matter phenomenon on the
cosmological scale without introducing other components in the theory.  At the
same time, as shown in \cite{SSV}, a low-density braneworld {\em with induced
gravity\/} can mimic the expansion properties of the LCDM model. In particular,
a universe consisting solely of baryons with $\Omega_{\rm b} \simeq 0.04$ can
mimic the LCDM cosmology with a much larger `effective' value of the matter
density $\Omega_{\rm m} \simeq 0.2$$-$$0.3$. This effect becomes possible owing
to the presence of a fundamental cross-over length scale $\ell$ in braneworld
models with induced gravity, which is absent in the Randall--Sundrum braneworld
model.

The aim of this paper is to investigate the properties of the braneworld model
with induced gravity to clarify further the role of the scale $\ell$ in the
possible explanation of the dark-matter phenomenon in astrophysics as well as
in cosmology. From the general consideration \cite{vdvz,SSV}, one expects the
braneworld model to resemble the general relativity at the distances $r\lesssim
\sqrt[3]{r_g\ell^2}$, thus giving an approximate Schwarzschild metric in vacuum
with $r_g$ in a role of the Schwarzschild radius. At large distances, the
modified gravitational field equations can be used to explain the properties of
the galactic rotation curves.

Observations of the 21~cm line from neutral hydrogen clouds in spiral galaxies
reveal that the rotational velocities remain nearly constant in the halo region
at values $\upsilon_c\sim 10^{-3}$$-$$10^{-4}$ in the units of the speed of
light. Starting from this observational fact, the exact metric in a galactic
halo was reconstructed in frames of the Randall--Sundrum braneworld model in
\cite{HC,BH}, which was used to account for this phenomenon by the effects of
higher-dimensional gravity without dark matter. We extend the results obtained
in these works to the case of braneworld model with induced gravity. Using the
smallness of $\upsilon_c$, we develop a perturbative scheme for reconstructing
metric in a galactic halo and demonstrate that, in the leading order of
expansion, at the distances $r\gtrsim \upsilon_c\ell$, our result reproduces
that obtained for the Randall--Sundrum braneworld model in \cite{HC,BH}. The
gravitational situation in this case can be approximately described by the
Einstein equations with the traceless projection of the bulk Weyl tensor to the
brane playing the role of an effective stress--energy tensor of galactic fluid.
To give a prescription for verifying the scenario based on the braneworld
models, we apply the results of \cite{FV}, in which it was proposed to use the
existing data on rotation curves and lensing measurements to constrain the
equation of state of the galactic fluid.

To solve the braneworld equations at small distances to the gravitating source,
one should impose some additional conditions. As first noted in \cite{SMS}, the
nonlocality and nonclosure of the braneworld equations are connected with the
projection $C_{ab}$ of the bulk Weyl tensor to the brane. It, therefore, seems
reasonable to impose certain restrictions on this tensor in order to obtain a
closed system of equations on the brane. In this way, several classes of vacuum
static spherically symmetric solutions on the brane were obtained for the
Randall--Sundrum braneworld model \cite{GM,HM}. Solutions obtained by imposing
direct restrictions on the form of the metric (such as $g_{tt}=-g_{rr}^{-1}$
\cite{bh}) also correspond to some evident conditions on $C_{ab}$. Finding the
correct boundary condition of this kind, one can turn the braneworld model into
a complete nonlinear {\em local\/} theory of gravity viable in all physical
circumstances.\footnote{Another possibilities to restrict the space of
solutions of braneworld equations in a spherically symmetric case was discussed
in \cite{HM,O}.}

The simplest way of restricting the tensor $C_{ab}$ consists in setting to zero
its (appropriately defined) anisotropic stress.\footnote{Earlier attempts to
simplify the braneworld equations by entirely neglecting the contribution from
$C_{ab}$ proved to be incorrect since this condition is incompatible with the
equation that follows from the Bianchi identity [see Eq.~\eqref{conserv}
below]. It was first noted in \cite{SMS} in the context of the Randall--Sundrum
model that neglecting $C_{ab}$ in the spherically symmetric situation leads to
a severe restriction on the energy-momentum tensor of perfect fluid, namely, it
only admits solutions with constant matter density (incompressible fluid).
Below, we demonstrate that this result is valid also in our more general
braneworld model with induced gravity.} This condition is fully compatible with
all equations of the theory and leads to a brane universe described by a
modified theory of gravity with an additional invisible component (Weyl fluid)
having nontrivial dynamics. A one-parameter family of boundary conditions (with
zero anisotropic stress as a particular case) was successfully employed in
\cite{SVS} to derive an exact system of equations describing scalar
cosmological perturbations on a generic braneworld with induced gravity.

In the present paper, we derive a closed system of equations for
static\footnote{For a discussion of the gravitational collapse on the brane see
\cite{gravitational collapse}.} spherically symmetric situation in the case of
zero anisotropic stress in the tensor $C_{ab}$ (the {\em minimal boundary
condition\/} in the terminology of \cite{SVS}). We find the Schwarzschild
metric to be an approximate vacuum solution of these equations at distances
$r\lesssim \sqrt[3]{r_g\ell^2}$, which justifies using of the minimal boundary
condition in the vicinity of gravitating objects.

The paper is organized as follows. In the next section, we remind the field
equations of the braneworld model with induced gravity. In
Sec.~\ref{sec:properties}, we discuss the generic properties of the braneworld
equations. In Sec.~\ref{sec:mimic}, we describe the cosmic mimicry model. In
Sec.~\ref{sec:rec}, we reconstruct the metric in a galactic halo with flat
rotation curves.  In Sec.~\ref{sec:bc}, we derive and discuss the system of
equations for the static spherically symmetric case with the minimal conditions
for the Weyl fluid. In Sec.~\ref{sec:vacuum}, we find vacuum solutions in the
weak field limit. In the last section, we present our conclusion.


\section{Effective field equations}
\label{sec:equations}

We start from the simplest generic braneworld model described by
the action
\begin{equation} \label{action}
S =  M^3 \left[\int_{\rm bulk} \left( {\cal R} - 2 \Lambda \right) -
2 \int_{\rm brane} K \right] + \int_{\rm brane} \left( m^2 R - 2
\sigma \right) + \int_{\rm brane} L \left( h_{ab}, \phi \right) \, .
\end{equation}
Here, ${\cal R}$ is the scalar curvature of the metric $g_{ab}$ in
the five-dimensional bulk, and $R$ is the scalar curvature of the
induced metric $h_{ab} = g_{ab} - n_a n_b$ on the brane ($n^a$ is
the vector field of the inner unit normal to the brane) which is
assumed to be a boundary of the bulk space, and the notation and
conventions of \cite{Wald} are used. The quantity $K = h^{ab}
K_{ab}$ is the trace of the symmetric tensor of extrinsic
curvature $K_{ab}$ of the brane. The symbol $L (h_{ab}, \phi)$
denotes the Lagrangian density of the four-dimensional matter
fields $\phi$ whose dynamics is restricted to the brane so that
they interact only with the induced metric $h_{ab}$. All
integrations over the bulk and brane are taken with the
corresponding natural volume elements. The symbols $M$ and $m$
denote the five-dimensional and four-dimensional Planck masses,
respectively, $\Lambda$ is the bulk cosmological constant, and
$\sigma$ is the brane tension.

Action (\ref{action}) leads to the Einstein equation with
cosmological constant in the bulk:
\begin{equation} \label{bulk}
{\cal G}_{ab} + \Lambda g_{ab} = 0 \, ,
\end{equation}
while the field equation on the brane is
\begin{equation} \label{brane}
m^2 G_{ab} + \sigma h_{ab} =  T_{ab} + M^3 \left(K_{ab} - h_{ab} K
\right) \, ,
\end{equation}
where $ T_{ab}$ is the stress--energy tensor on the brane stemming from the
last term in (\ref{action}). The Einstein equation in the bulk \eqref{bulk} and
the Gauss--Codazzi identities imply the conservation of the stress--energy
tensor of matter on the brane, namely, $\nabla^a T_{ab}=0$ (see, for example,
\cite{Sh}). Here, $\nabla_a$ denotes the covariant derivative on the brane
associated with the induced metric $h_{ab}$.

By using the Gauss--Codazzi identities and projecting the field equations to
the brane, one can obtain the following effective equation \cite{SSV}:
\begin{equation}\label{effective}
G_{ab} + \frac{\Lambda_{\rm RS}}{\beta + 1} h_{ab} = \left(\frac{\beta}{\beta +
1}\right) \frac{1}{m^2}  T_{ab} + \frac{1}{\beta + 1} \left( \frac{1}{M^6}
Q_{ab} - C_{ab} \right) \, ,
\end{equation}
where
\begin{equation}\label{beta}
\beta = k \ell\,, \qquad k=\frac{\sigma}{3M^3}\,, \qquad \ell=\frac{2m^2}{M^3}
\end{equation}
are convenient parameters of the braneworld model,
\begin{equation} \label{lambda-eff}
\Lambda_{\rm RS}=\frac{\Lambda}{2}+\frac{\sigma^2}{3M^6}
\end{equation}
is the value of the effective cosmological constant in the Randall--Sundrum
model,
\begin{equation}\label{q}
Q_{ab} = \frac13 E E_{ab} - E_{ac} E^{c}{}_b + \frac12 \left(E_{cd}
E^{cd} - \frac13 E^2 \right) h_{ab}
\end{equation}
is a quadratic expression with respect to the `bare' Einstein equation $E_{ab}
\equiv m^2 G_{ab} -  T_{ab}$ on the brane, and $E = h^{ab} E_{ab}$. The
symmetric traceless tensor $C_{ab} \equiv n^c n^d C_{acbd}$ is a projection of
the bulk Weyl tensor $C_{abcd}$ which carries information about the
gravitational field outside the brane.  The tensor $C_{ab}$ is not freely
specifiable on the brane, but is related to the tensor $Q_{ab}$ through the
conservation equation
\begin{equation}\label{conserv}
\nabla^a \left( Q_{ab} - M^6 C_{ab} \right) = 0 \, ,
\end{equation}
which is a consequence of the Bianchi identity applied to \eqref{effective}.

The following famous braneworld theories are related to important subclasses of
action (\ref{action}):
\begin{enumerate}
\item The Randall--Sundrum model \cite{RS} is obtained after setting $m=0$ in
(\ref{action}). The corresponding effective equation on the brane was derived
in \cite{SMS}:
\begin{equation}\label{effective RS}
G_{ab} + \Lambda_{\rm RS} h_{ab} = \frac{2\sigma}{3M^6}  T_{ab} + \frac{1}{M^6}
S_{ab} - C_{ab} \, ,
\end{equation}
where
\begin{equation}\label{q RS}
S_{ab} = \frac13  T  T_{ab} -  T_{ac}  T^{c}{}_b + \frac12 \left(
T_{cd}  T^{cd} - \frac13  T^2 \right) h_{ab}\,.
\end{equation}

In this model, only the right-hand side of the Einstein equation is modified.

\item The Dvali--Gabadadze--Porrati (DGP) model \cite{DGP} corresponds to the special
case where both the cosmological constant in the bulk and the brane tension
vanish, i.e., $\Lambda = 0$ and $\sigma = 0$ in (\ref{action}). The
corresponding effective equation
\begin{equation}\label{effectiv DGP}
    G_{ab}=\frac{1}{M^6}Q_{ab}-C_{ab}
\end{equation}
does not contain a linear contribution from the stress--energy tensor $
T_{ab}$, which makes a perturbative analysis in this theory problematic.

\item Finally, general relativity, leading to the LCDM cosmological model, is
formally obtained after setting $M = 0$ in (\ref{action}). In this case, the
effective equation $Q_{ab}=0$ has an obvious solution
\begin{equation}\label{einstein limit}
    G_{ab}=\frac{1}{m^2} T_{ab}\,.
\end{equation}
\end{enumerate}


\section{Generic properties of braneworld gravity}
\label{sec:properties}

The theory described by action \eqref{action} has two important scales, defined
in \eqref{beta}, namely, the cross-over length scale $\ell$, which describes
the interplay between the bulk and brane gravity, and the mass (energy) scale
$k$, which determines the role of the brane tension in the dynamics of the
brane. Thus, we can expect the properties of braneworld gravity to be different
at different scales.

Indeed, looking at \eqref{effective}, we notice that expression \eqref{q} for
$Q_{ab}$ is quadratic in the curvature as well as in the stress--energy tensor.
Therefore, in the region of high matter density and curvature, the tensor
$Q_{ab}$ dominates in \eqref{effective}, and the gravitational law is
approximated by the `bare' Einstein equations $m^2 G_{ab}-T_{ab}=0$. And, vice
versa, the contribution from $Q_{ab}$ in \eqref{effective} is insignificant on
sufficiently large length scales, where the curvature is small, and the
braneworld theory on those scales should again be approximated by the Einstein
gravity with different gravitational constant. The tensor $C_{ab}$ in this case
plays the role of some effective stress--energy tensor.

To give some qualitative numerical estimates, we repeat here the reasoning of
\cite{SSV}. Consider the trace of the effective equation \eqref{effective}:
\begin{equation}\label{trace of effective}
-R + \frac{4\Lambda_{\rm RS}}{\beta + 1} -
\left(\frac{\beta}{\beta + 1}\right) \frac{1}{m^2}  T =
\frac{1}{\beta + 1} \frac{1}{M^6} Q \, ,
\end{equation}
where the left-hand side contains terms which are linear in the curvature and
in the stress--energy tensor while the right-hand side contains the quadratic
term $Q = h^{ab} Q_{ab}$. This equation is closed on the brane in the sense
that it does not contain the contribution from the traceless tensor $C_{ab}$.

Suppose that we are interested in the behaviour of gravity in the neighborhood
of a spherically symmetric object with density $\rho_s$, total mass ${\cal
M}_s$, and radius $r_s$. For simplicity, we assume that one can neglect the
tensor $C_{ab}$ and the effective cosmological constant in the neighborhood of
the source. As regards the effective cosmological constant, this assumption is
natural. Concerning the neglect of the tensor $C_{ab}$, we demonstrate the
validity of this approximation in Sec.~\ref{sec:vacuum} by considering the
vacuum spherically symmetric situation in the braneworld model with $C_{ab}$
restricted by a simple condition of vanishing anisotropic stress.

Within the source itself, we have two qualitatively different options: an
approximate solution can be sought either neglecting the quadratic part or
linear part of \eqref{effective} and \eqref{trace of effective}. We should
choose the option that gives smaller error of approximation in \eqref{trace of
effective}. In the first case, neglecting the quadratic part and the effective
cosmological constant, we have
\begin{equation}\label{lin}
G_{ab} - \left(\frac{\beta}{\beta + 1}\right) \frac{1}{m^2} T_{ab} \approx 0
\quad \Rightarrow \quad \frac{Q}{(\beta + 1) M^6} \sim \frac{\rho_s^2}{(\beta +
1)^3 M^6} \, .
\end{equation}
In the second case, we neglect the linear part, so that
\begin{equation}\label{quadrat}
Q_{ab} \approx 0 \quad \Rightarrow \quad E_{ab} \approx 0 \quad \Rightarrow
\quad  R + \left(\frac{\beta}{\beta + 1}\right) \frac{1}{m^2} T \sim
\frac{\rho_s}{(\beta + 1) m^2 } \, .
\end{equation}
The final expression on the right-hand side of \eqref{quadrat} is
smaller than the corresponding expression in \eqref{lin} if
\begin{equation}\label{scale}
\rho_s > (\beta + 1)^2 \frac{M^6}{m^2} \quad \Rightarrow \quad r_s^3 < r_*^3
\sim \frac{{\cal M}_s \ell^2}{(\beta + 1)^2 m^2 } \, ,
\end{equation}
where we used the relation ${\cal M}_s \sim \rho_s r_s^3$. Thus, we can expect
that, in the neighborhood of the source, on distances smaller than $r_*$ given
by \eqref{scale}, the solution is determined mainly by the quadratic part
$Q_{ab}$ in \eqref{effective}, which means that it respects the `bare' Einstein
equation $m^2 G_{ab} - T_{ab}=0$ to a high precision. This effect is sometimes
described as the `gravity filter' of the DGP model \cite{DGP}, which screens
the scalar graviton in the neighborhood of the source making the gravity
effectively Einsteinian.  Some aspects of this interesting phenomenon were
discussed in \cite{Kaloper}.

Expression \eqref{scale} generalizes the length scale \cite{vdvz}
of the DGP model, below which nonlinear effects become important,
to the case of nonzero brane tension (nonzero $\beta$) and bulk
cosmological constant satisfying the Randall--Sundrum constraint
$\Lambda_{\rm RS}=0$. In order to comply with observations in the
solar system, the value of $\ell$ should be chosen sufficiently
large.  In the next section, we will see that modeling the
homogeneous dark matter phenomenon by cosmic mimicry requires
$\beta \approx - 5/4$, and in Sec.~\ref{sec:rec} we will show that
matching the galactic rotation curves can be realized if $\ell
\lesssim 2$~Mpc. For these parameter values, the value of $r_*$
for the Sun is of the order of $1$~pc, which is quite large so
that the solar-system experiments are in accord with the
general-relativistic expectations.

\section{Cosmic mimicry on the brane}
\label{sec:mimic}

In our paper \cite{SSV}, we described a braneworld model in which cosmological
evolution proceeds similarly to that of the Friedmannian cosmology but with
different values of the effective matter parameter $\Omega_{\rm m}$, or,
equivalently, with different values of the effective gravitational constant
$G_{\rm N}$, at different cosmological epochs.

Specifically, for a spatially flat universe with zero background dark radiation
($C = 0$), the cosmological equation stemming from (\ref{action}) can be
written as follows:
\begin{equation} \label{mimic}
H^2 = \frac{\Lambda}{6} + \left[ \sqrt{ \frac{\rho - \rho_0}{3 m^2} + \left(
\sqrt{H_0^2 - \frac{\Lambda}{6}} \mp \frac1\ell \right)^2 } \pm \frac1\ell
\right]^2 \, ,
\end{equation}
where $\rho_0$ and $H_0$ are the energy density of matter and Hubble parameter,
respectively, at the present moment of time.  The two signs in (\ref{mimic})
correspond to two complementary possibilities for embedding the brane in the
higher-dimensional (Schwarzschild-AdS) bulk.

In the effective mimicry scenario \cite{SSV}, the parameters $H_0^2 - \Lambda /
6$ and $1 / \ell^2$ are assumed to be of the same order, and much larger than
the present matter-density term $\rho_0 / 3 m^2$. The mimicry model has two
regimes: one in the deep past (where the matter density was high) and another
during the late-time evolution (where the matter density is low). In the deep
past, we have
\begin{equation}
\frac{\rho - \rho_0}{3 m^2} \gg \left(\sqrt{H_0^2 - \frac{\Lambda}{6}} \mp
\frac{1}{\ell} \right)^2 \, ,
\end{equation}
and the universe expands in a Friedmannian way
\begin{equation} \label{past}
H^2 \approx \frac{\rho}{3 m^2} \, .
\end{equation}

During the late-time evolution, we have
\begin{equation}
\frac{\rho - \rho_0}{3 m^2} \ll \left(\sqrt{H_0^2 - \frac{\Lambda}{6}} \mp
\frac{1}{\ell} \right)^2 \, ,
\end{equation}
and the expansion law is approximated by
\begin{equation} \label{future}
H^2 \approx H_0^2 + \frac{\alpha}{\alpha \mp 1} \frac{\rho - \rho_0}{3 m^2} \,
,
\end{equation}
where
\begin{equation}\label{alpha}
\alpha = \ell \sqrt{H_0^2 - \frac{\Lambda}{6}}
\end{equation}
is the parameter introduced in \cite{SSV}.  In the case of the cosmological
branch with upper sign, it is assumed that the coefficient $\alpha / (\alpha -
1)$ in (\ref{future}) and in similar expressions is always positive, i.e.,
$\alpha$ is assumed to be greater than unity in this case.

One can interpret the result (\ref{future}) either as a renormalization of the
effective gravitational constant relative to its value in the deep past or as a
renormalization of the effective density parameter:
\begin{equation} \label{lcdm}
H^2 \approx H_0^2 + \frac{\rho^{\rm LCDM} - \rho^{\rm LCDM}_0}{3 m^2} \, ,
\qquad \rho^{\rm LCDM} = \frac{\alpha}{\alpha \mp 1} \rho \, .
\end{equation}

Remarkably, the behaviour of the Hubble parameter on the brane practically {\em
coincides\/} with that in LCDM at low densities. This property was called {\em
`cosmic mimicry'\/} in \cite{SSV}. A consequence of this is the fact that a
low/high density braneworld consisting entirely of baryonic matter with
$\Omega_{\rm b} \simeq 0.04$ could easily masquerade as LCDM with a moderate
value $\Omega^{\rm LCDM}_{\rm m} \simeq 0.2$$-$$0.3$. Thus, with the upper sign
in (\ref{lcdm}), we can explain the phenomenon of homogeneous dark matter if
suppose $\alpha\approx 5/4$. In this case, taking into account that the value
of $\ell$ lies well below the Hubble scale, from \eqref{alpha} we obtain
\begin{equation}\label{estimation of lambda}
    |\Lambda|\ell^2 \approx 6\alpha^2 \approx 9 \,.
\end{equation}
The value of $\Lambda$ should be negative for the expression under
the square root in \eqref{alpha} to be positive. As was shown in
\cite{SSV}, $\beta$ is also negative in this case, and
$|\beta|\approx\alpha$. The upper estimate $\ell \lesssim 2$~Mpc
is obtained in the next section by considering galactic rotation
curves. The range of redshifts over which this cosmic mimicry
occurs is given by $0 \leq z \ll z_{\rm m}$, where $z_{\rm m}$ in
this case can be estimated to be $z_{\rm m} \gtrsim 170$.


\section{Reconstruction of the metric in a galactic halo}
\label{sec:rec}

In the previous section, we described how the phenomenon of homogeneous dark
matter can be explained by cosmic mimicry in the braneworlds models with
induced gravity. The question then arises whether the phenomenon of dark matter
can also be accounted for in the inhomogeneous situations, in particular, on
the scales of galaxies.  This is the subject of the present section.

The problem of reconstructing the static metric and dark-matter halos in
general relativity from the behavior of galactic rotation curves was
investigated recently in \cite{NSS,BK,FV}.  A similar procedure was applied to
the Randall--Sundrum braneworld model in \cite{PBK,HC,BH}, with a conclusion
that it can explain the observations of the galactic rotation curves without
dark matter.  The Randall--Sundrum model, however, does not allow for the
mimicry property and thus cannot explain the {\em homogeneous\/} dark-matter
phenomenon. Hence, it is necessary to address this issue in frames of the
induced-gravity model.

The traceless tensor $C_{ab}$ drops out completely from the trace of
\eqref{effective}, which is the only closed equation on the brane in the
absence of any additional constraints on the tensor $C_{ab}$ \cite{Shtanov}.
In the vacuum, the trace of \eqref{effective} reads
\begin{equation} \label{trace}
(1 + \beta) R + \frac14 \ell^2 \left( R_{ab} R^{ab} - \frac13 R^2 \right) - 4
\Lambda_{\rm RS} = 0 \, .
\end{equation}
One can use this equation to reconstruct the metric in the neighborhood of a
galaxy by assuming approximate spherical symmetry and taking into account the
qualitative behavior of rotation curves.  In the Randall--Sundrum model [which
corresponds to setting $\ell = 0$ and $\beta = 0$ in \eqref{trace}], this
procedure was done in \cite{PBK,HC,BH}.

The spherically symmetric metric, by which we approximate the situation in a
galactic halo, has the form
\begin{equation}\label{halo-metric}
    ds^2=-f(r)dt^2+ g^{-1} (r)dr^2+r^2(d\theta^2+\sin^2\theta
    d\varphi^2)\,.
\end{equation}
In the region of the flat rotation curve, we have the condition
\begin{equation}
    \phi'(r)=\frac{\upsilon^2_c}{r} \, ,
\end{equation}
where $\upsilon_c$ is the rotation velocity, and $\phi (r) = \frac12 \log
f(r)$. It can be integrated to give the $g_{00}$ component of the metric:
\begin{equation} \label{f}
f (r) = \left( \frac{r}{r_c} \right)^{2 \upsilon_c^2} \, ,
\end{equation}
where $r_c$ is the constant of integration.  Substituting metric
\eqref{halo-metric} with the function $\phi (r)$ given by
\eqref{f} into equation \eqref{trace}, we obtain the following
first-order nonlinear differential equation for the function
$g(r)\,$:
\begin{eqnarray} \label{g}
&{}& \left(rg'-2g\right)^2 + 4rg'\left(1-\frac{12(1+\beta)r^2}{\ell^2}\right)
-8g\left(1+\frac{6(1+\beta)r^2}{\ell^2}\right) \nonumber \\
&{}& {} + 4\left[1+\frac{12r^2\left(1+\beta-2\Lambda_{\rm RS}r^2\right)}{\ell^2}\right] \nonumber \\
&{}&
{}+\upsilon_c^2\left[-2r^2(g')^2-24rgg'+8g^2+8rg'
\left(1-\frac{3(1+\beta)r^2}{\ell^2}\right)-8g\left(1+\frac{6(1+\beta)r^2}{\ell^2}\right)\right]
\nonumber \\
&{}& {}+ \upsilon_c^4\left[(g')^2-12rgg'+16g
\left(1-\frac{3(1+\beta)r^2}{\ell^2}\right)\right]
+4\upsilon_c^6g\left(rg'-4g\right)+4\upsilon_c^8g^2=0
\, .
\end{eqnarray}

This equation looks rather complicated. However, we can take into account that
the quantity $\upsilon_c^2 \sim 10^{-7}$ is very small for typical galaxies and
thus represents a convenient parameter for asymptotic expansion. Therefore, we
can look for the solution of \eqref{g} in the form of expansion in powers of
$\upsilon_c^2\,$:
\begin{equation}\label{asympt g}
    g(r)= g_0(r)+\upsilon_c^2 g_1(r)+\upsilon_c^4
    g_2(r)+ \ldots \,,
\end{equation}
where the functions $g_0$, $g_1$, $g_2$, $\ldots$, should be determined step by
step by solving \eqref{g} with accuracy corresponding to the order of
$\upsilon_c$ at every step. First, we should find $g_0(r)$, which is the
solution of \eqref{g} with all powers of $\upsilon_c$ neglected. Introducing
$\chi(r)=rg_0(r)$, we can rewrite the equation for $g_0(r)$ in the following
form:
\begin{equation}\label{y}
   \ell^2 \left(\chi'-\frac{3\chi}{r}+2\right)^2+48 r^2 \left[(1+\beta)(1-\chi')-2\Lambda_{\rm
   RS}r^2\right]=0\,.
\end{equation}

In the Randall--Sundrum limit, which implies $\ell=0$ and $\beta=0$, we should
have $\chi_1(r)=r + a - 2\Lambda_{\rm RS}r^3/3$ as a solution of \eqref{y},
where $a$ is some arbitrary constant of integration. On the other hand, the
function $\chi_2(r)=r+ b r^3$, with another arbitrary constant $b$, sets
expression in the first parentheses to zero. One can easily observe that the
function $\chi(r)=r-2\Lambda_{\rm RS}r^3/3(1+\beta)$ is an exact particular
solution of \eqref{y}, setting to zero both components on the left-hand side
separately. Thus, for the zero-order approximation, we have
\begin{equation}\label{g0}
    g_0(r)=1-\frac{2\Lambda_{\rm RS}r^2}{3(1+\beta)}\,.
\end{equation}

The expression $2\Lambda_{\rm RS}/(1+\beta)$ here represents an analog of the
cosmological constant in general relativity. At the scales at which galactic
rotation curves are observed, the contribution from the cosmological constant
should be negligibly small, namely
\begin{equation}\label{Lambda r}
     \frac{\Lambda_{\rm RS}r^2}{|1+\beta|}\ll 1\,,
\end{equation}
thus allowing us to neglect it in the future. We should keep it for a moment,
to estimate corrections in the second order of $\upsilon_c$.

Using \eqref{asympt g} and \eqref{g0}, we find the differential equation for
$g_1(r)\,$:
\begin{equation}\label{dif g1}
    rg_1'+g_1=-1+\frac{\Lambda_{\rm
    RS}\ell^2}{3(1+\beta)^2}\left[1-\frac{4\Lambda_{\rm
    RS}r^2}{3(1+\beta)}\left(1-\frac{3(1+\beta)^2}{\Lambda_{\rm
    RS}\ell^2}\right)\right]\,,
\end{equation}
which can easily be integrated to give
\begin{equation}\label{g1}
    g_1(r)=-\left(1+\frac{C_1}{r}\right)+\frac{\Lambda_{\rm
    RS}\ell^2}{3(1+\beta)^2}\left[1-\frac{4\Lambda_{\rm
    RS}r^2}{9(1+\beta)}\left(1-\frac{3(1+\beta)^2}{\Lambda_{\rm
    RS}\ell^2}\right)\right]\,,
\end{equation}
where $C_1$ is some constant of integration having the dimension of length.

In the cosmic-mimicry theory which, as described in the previous
section, explains the dark-matter phenomenon on the cosmological
scale, the value of $\ell$ lies well below the cosmological length
scale. Thus, we have
\begin{equation}\label{Lambda ell}
     \frac{\Lambda_{\rm
    RS}\ell^2}{(1+\beta)^2}\ll 1 \, .
\end{equation}
Conditions \eqref{Lambda r} and \eqref{Lambda ell} allow us to write the
function $g(r)$ at this stage in the form
\begin{equation}\label{g up to second order}
    g(r)\approx
    1-\frac{2\Lambda_{\rm RS}r^2}{3(1+\beta)}-\upsilon_c^2
    \left(1+\frac{C_1}{r}\right)\,.
\end{equation}

It is important to note that, under conditions \eqref{Lambda r} and
\eqref{Lambda ell}, the difference between $g(r)$ in our model and the same
quantity in the Randall--Sundrum braneworld model in this approximation in
$\upsilon_c^2$  is only in the appearance of the factor $(1+\beta)$, which
simply renormalizes the cosmological constant.

To find corrections of the forth order in $\upsilon_c$, we shall completely
neglect the contribution from the term containing $\Lambda_{\rm RS}$. This can
be done under the condition
\begin{equation}\label{Lambda r v}
    r^2\ll \frac{\upsilon_c^2|1+\beta|}{\Lambda_{\rm RS}}\,,
\end{equation}
which, obviously, is more restrictive than \eqref{Lambda r}, but still is
satisfied for a galaxy. The result is
\begin{equation}\label{g v4}
    g(r)\approx
    1-\upsilon_c^2-\frac{\upsilon_c^2 C_1}{r}
    \left[1-\frac{\upsilon_c^2}{2}\ln\frac{r}{\left|C_1\right|}\right]
    -\frac{\ell^2\upsilon_c^4}{4(1+\beta)r^2}\left(1 - \frac{C_1}{2r} +
    \frac{C_1^2}{4r^2} \right)\,,
\end{equation}
in which the constant of integration $C_1$ is renormalized as compared with the
previous approximation.

Although the last term in \eqref{g v4} is proportional to
$\upsilon_c^4$, at too small distances it can turn out to be
larger than $\upsilon_c^2$. This will obviously destroy our
perturbative analysis. To avoid such situation, we restrict
ourselves to the region\footnote{For simplicity, we assume the
length scale $|C_1|$ to satisfy the condition
$|C_1|\lesssim\upsilon_c \ell/\sqrt{|1+\beta|}$, so that $\left|
C_1 \right| / r$ is small. However, we would like to note here an
interesting possibility that a large constant $|C_1|$ determines
another interesting scale in a galactic halo.}
$r^2\gg\ell^2\upsilon_c^2/|1+\beta|$.

Leaving only corrections of the leading order in \eqref{g v4}, we can write
\begin{equation}\label{g approx}
    g(r)\approx
    1-\upsilon_c^2\left(1+\frac{C_1}{r}\right)\,,
\end{equation}
which is a general reconstruction of the metric in our braneworld model in the
region
\begin{equation}\label{condition of pert}
     \frac{\ell^2\upsilon_c^2}{|1+\beta|}\ll r^2 \ll \frac{\upsilon_c^2|1+\beta|}{\Lambda_{\rm RS}}\,
\end{equation}
with condition \eqref{f}.

In order to satisfy the left inequality in \eqref{condition of pert} in a real
spiral galaxy such as our Milky Way with the distance $ r \sim 3$~kpc at which
the rotation velocity becomes constant of magnitude $\upsilon_c \sim 7\times
10^{-4}$, we should have $\ell \lesssim \sqrt{|1 + \beta|}\, \upsilon_c^{-1}\,
r \simeq 2$~Mpc, where we took into account the mimicry estimate $\beta = -
\alpha \approx - 5/4$ (see the end of Sec.~\ref{sec:mimic}).  In this paper, we
restrict our investigation to this case.  For magnitudes of the cross-over
length scale $\ell$ essentially larger than this value, one cannot use the
perturbation expansion in powers of $\upsilon_c^2$, and it is necessary to
solve the nonlinear equation \eqref{g}.

We can substitute \eqref{f}, \eqref{g approx} into \eqref{effective} to compute
the tensor $C_{ab}$. The result is
\begin{equation}\label{Cab}
    C^t{}_{t}\approx \frac{(1+\beta)\upsilon_c^2}{r^2}\,, \quad C^r{}_{r}\approx
    \frac{(1+\beta)\upsilon_c^2}{r^2}\left(\frac{C_1}{r}-1\right)\,, \quad
    C^\theta{}_{\theta}=C^\varphi{}_{\varphi}\approx
    -\frac{(1+\beta)\upsilon_c^2 C_1}{2r^3}\,,
\end{equation}
where we have neglected the contribution from terms of order
$\upsilon_c^4$, which is correct in region \eqref{condition of
pert}. Again, we observe that, in the leading order of expansion,
our result reproduces that obtained for the Randall--Sundrum
braneworld model in \cite{HC,BH}. This is what we have expected
from general consideration of Sec.~\ref{sec:properties}: on
sufficiently large length scales, the contribution from the
quadratic expression $Q_{ab}$ in \eqref{effective} is negligibly
small. The gravitational situation in this case can be
approximately described by the Einstein equations with the tensor
$C_{ab}$ playing the role of some effective stress--energy tensor
$\widetilde{T}_{ab}$:
\begin{equation}\label{effectime without Q}
    8\pi G_{\rm N}
    \widetilde{T}_{ab}\equiv -\frac{1}{1+\beta}C_{ab}=G_{ab} \,,
\end{equation}
where $G_{\rm N}$ is the gravitational constant adopted by an observer on these
scales.

Using \eqref{Cab} and \eqref{effectime without Q}, we determine the effective
energy density $\widetilde{\rho}(r)$ and radial and transverse pressures
$\widetilde{p}_r(r)$ and $\widetilde{p}_t(r)$, respectively, inside the
galactic halo:
\begin{equation}\label{halo fluid}
8\pi G_{\rm N}\widetilde{\rho}(r)\approx \frac{\upsilon_c^2}{r^2}\,, \qquad
8\pi G_{\rm N}\widetilde{p}_r(r)\approx
\frac{\upsilon_c^2}{r^2}\left(1-\frac{C_1}{r}\right)\,, \qquad 8\pi G_{\rm
N}\widetilde{p}_t(r)\approx \frac{\upsilon_c^2}{r^2}\, \frac{C_1}{2r}\,.
\end{equation}
We note that $\widetilde{p}_r(r)\neq\widetilde{p}_t(r)$, which indicates the
presence of a significant anisotropic stress in the tensor $C_{ab}$ [see
definition \eqref{weil definition} below] on galactic scales.

It was argued in \cite{FV} that combined observations of galactic rotation
curves and gravitational lensing can yield the profile of pressure inside a
galactic halo. The authors of \cite{FV} proposed to use the pseudo-masses
$m_{\rm RC}(r)$ and $m_{\rm lens}(r)$, which can be obtained for the same
galactic halo from the treatment of rotation curves and gravitational lensing
in the context of dark-matter paradigm (which implies negligible pressure), to
define the $\chi$-factor which quantifies the deviation from the predictions of
the CDM model:
\begin{equation}\label{chi}
\chi[\omega(r)] = \frac{m'_{\rm lens}(r)}{m'_{\rm RC}(r)} =
\frac{2+3\omega(r)}{2+6\omega(r)}\,,
\end{equation}
where
\begin{equation}\label{omega}
    \omega(r)=\frac{\widetilde{p}_r(r)+2\widetilde{p}_t(r)}{3\widetilde{\rho}(r)}\,.
\end{equation}

For the braneworld model, using \eqref{halo fluid}, one obtains $\omega\approx
1/3$ and $\chi\approx 3/4$, meaning that deflection angles, which can be
computed in our model, will be around $75\%$ of the usual predictions based on
dark matter. This result coincides with the analysis made in \cite{PBK}.

\section{Minimal condition for the Weyl fluid on small scales}
\label{sec:bc}

In the previous section, we have shown that the contribution of $Q_{ab}$ to
\eqref{effective} is negligibly small on sufficiently large spatial scales,
which makes it possible to explain the flatness of galactic rotation curves as
an effect of higher-dimensional gravity.  On such scales, the braneworld theory
reduces to the Einstein gravity with the tensor $C_{ab}$ playing the role of
some effective stress--energy tensor.

To demonstrate that the contribution from $C_{ab}$ to the gravitational
dynamics is insignificant at small spatial scales, we should solve
Eq.~\eqref{effective} at small distances from the source of gravity. The
problem arising here is that equation (\ref{effective}) is not closed on the
brane in the sense that the dynamics of the symmetric traceless tensor $C_{ab}$
on the brane is not determined by the dynamics of matter alone. An
approximation usually taken by several authors to overcome this problem
consists in imposing certain conditions on the tensor $C_{ab}$ directly on the
brane so as to close Eq.~(\ref{effective}). Such conditions on the tensor
$C_{ab}$ should be compatible with the conservation equation (\ref{conserv})
and should also leave the braneworld theory compatible with Einstein's general
relativity in a wide range of physical situations.

In general, the tensor $C_{ab}$ can be decomposed through an arbitrary
normalized timelike vector field $u^a$ on the brane (see \cite{M}):
\begin{equation}\label{weil definition}
m^2 C_{ab} = \frac{\rho_{\cal C}}{3} \left(h_{ab}+4u_a
u_b\right)+2\upsilon^{\cal C}_{(a}u^{}_{b)} + \pi^{\cal C}_{ab}\,
.
\end{equation}
Here, the covector $\upsilon^{\cal C}_{a}$ and traceless symmetric tensor
$\pi^{\cal C}_{ab}$ are both orthogonal to $u^a$.  The tensor $C_{ab}$ in this
case is regarded as the stress--energy tensor of an ideal fluid with equation
of state like that for radiation but with nontrivial dynamics described by
Eq.~(\ref{conserv}). In the literature, it is called `Weyl fluid'
\cite{W-fluid} and, in the cosmological context, `dark radiation'
\cite{dark-rad}. The quantity $\rho_{\cal C}$ has the meaning of its density,
$\upsilon^{\cal C}_a$ is its momentum transfer, and $\pi^{\cal C}_{ab}$ is its
anisotropic stress, as measured by observers following the world lines tangent
to $u^a$.  The stress--energy of this ideal fluid is not conserved due to the
presence of the source term $Q_{ab}$ in (\ref{conserv}).

Equation (\ref{conserv}) gives evolution equations for the components
$\rho_{\cal C}$ and $\upsilon_{\cal C}^a$.  However, there are no evolution
equations for the tensor fields $\pi^{\cal C}_{ab}$ on the brane, which is a
manifestation of the nonlocality of the physical situation from the braneworld
viewpoint. Thus, boundary conditions for the brane--bulk system can be
specified by imposing additional conditions on the tensor $C_{ab}$.

As the first condition on the tensor field $C_{ab}$, we demand that it has a
normalized timelike eigenvector field $u^a_{\cal C}$, so that $C^a{}_b
u^b_{\cal C} \propto u^a_{\cal C}$.  Then this vector field can be used as
vector $u^a$ in decomposition \eqref{weil definition}, which was not specified
up to now:
\begin{equation}\label{weil definition1}
m^2 C_{ab} = \frac{\rho_{\cal C}}{3} \left(h_{ab}+4u_a^{\cal C} u_b^{\cal C}
\right) + \pi^{\cal C}_{ab}\, .
\end{equation}
Note that the covector component $v_a^{\cal C}$ vanishes in this case because
of the eigenvector property of $u^a_{\cal C}$.  The Weyl fluid is now described
by the timelike eigenvector field $u^a_{\cal C}$, which can be interpreted as
its four-velocity, and by the quantity $\rho_{\cal C}$, which has a meaning of
its rest-frame density.

Now, in solving the spherically symmetric problem in the neighborhood of the
source, we assume that the anisotropic stress can be neglected in the natural
decomposition \eqref{weil definition1}:
\begin{equation} \label{bc}
\pi^{\cal C}_{ab} = 0 \, .
\end{equation}
This is a {\em minimal boundary condition\/} for the brane--bulk system in the
terminology of \cite{SVS}. Under condition \eqref{bc}, the normalized
eigenvector field $u^a_{\cal C}$ is unique modulo orientation.

Equation (\ref{conserv}) gives the evolution equations for the
components $\rho_{\cal C}$ and $u_{\cal C}^a$.  Hence,
\eqref{effective}, together with equations for material fields,
leads to a complete set of equations.

In what follows, we assume \eqref{bc} to be the case at small distances to the
source of gravity and solve the vacuum static spherically symmetric problem
with the additional condition \eqref{bc}. On large scales, this conditions is
violated [see Eq.~\eqref{halo fluid}].

A general static spherically symmetric metric on the brane has the form
\begin{equation}\label{spherically symmetric metric}
    ds^2=-f(r)dt^2+h(r)dr^2+r^2(d\theta^2+\sin^2\theta
    d\varphi^2)\,.
\end{equation}
It leads to the following nonzero components of the Einstein tensor:
\begin{equation}\label{G00}
    - G^t{}_t=\frac{h'}{rh^2}+\frac{1}{r^2} \left( 1-\frac{1}{h} \right)\,,
\end{equation}
\begin{equation}\label{G11}
    G^r{}_r=\frac{f'}{rfh}-\frac{1}{r^2} \left( 1-\frac{1}{h} \right)\,,
\end{equation}
\begin{equation}\label{G22}
G^\theta{}_\theta = G^\varphi{}_\varphi = \frac{f''}{2fh} - \frac{f'h'}{4fh^2}
- \frac{f'^2}{4f^2h} + \frac{f'}{2rfh} - \frac{h'}{2rh^2}\,.
\end{equation}

The interior of static objects (such as stars) is naturally described by the
stress--energy tensor of perfect fluid $ T_{ab}=\rho u_a u_b +p(h_{ab}+u_a
u_b)$. To be compatible with the space-time symmetry, the fluid four-velocity
$u^a$ must be aligned with the static Killing vector field $\xi^a\,$:
\begin{equation}\label{fluid velocity}
    u^a=-\sqrt{f} \left(\frac{\partial}{\partial t} \right)^a\,,
\end{equation}
so that matter is described by the two functions $\rho(r)$ and $p(r)$, and the
coordinate components of its stress--energy tensor are given by
\begin{equation} \label{tau}
 T^\alpha{}_\beta = \left(
\begin{array}{rl}
 {-\rho(r)}
\, , &  \overrightarrow{0} \medskip \\
\overrightarrow{0} \, , & p(r)\delta^i{}_j
\end{array}
\right) \, .
\end{equation}

The spherical symmetry and static property imply that the Weyl fluid
four-velocity is also aligned with the static Killing vector
field:\footnote{The four-velocity of the Weyl fluid certainly will play an
important role in a non-static situation.}
\begin{equation}
    u^a_{\cal C}=-\sqrt{f}\left(\frac{\partial}{\partial t} \right)^a\,,
\end{equation}
so that the Weyl fluid is described by the single function $\rho_{\cal C}(r)$:
\begin{equation} \label{weyl}
m^2 C^\alpha{}_\beta = \left(
\begin{array}{rl}
 {-\rho_{\cal C}(r)}
\, , &  \overrightarrow{0} \medskip \\
\overrightarrow{0} \, , & \frac{1}{3}\rho_{\cal C}(r) \delta^i{}_j
\end{array}
\right) \, .
\end{equation}
One should note that $C^r{}_r=C^\theta{}_\theta$ as a result of our boundary
condition \eqref{bc} which requires vanishing of the anisotropic stress. The
tensor $C_{ab}$ manifests itself directly as dark radiation in the
Randall--Sundrum model [see Eq.~\eqref{effective RS}]; however, in a braneworld
model with induced gravity, which is described by the effective equation
\eqref{effective}, its influence on the gravitational dynamics is more
complicated, and new effects can be expected.

To proceed further, we rewrite Eq.~\eqref{effective} as
\begin{equation}\label{effective rewritten}
 Q_{ab}=\frac{M^6(1+\beta)}{m^2}E_{ab}+M^6\left(\frac{1}{m^2} T_{ab}+C_{ab}+\Lambda_{\rm RS}h_{ab}\right)\,.
\end{equation}
Here, $Q_{ab}$ is the quadratic expression with respect to the nonzero diagonal
components of $E_{ab}$ given by \eqref{q}. Introducing
\begin{equation}\label{definition xyz via E}
   x= - E^t{}_t\,, \qquad y=E^r{}_r\,, \qquad z=E^\theta{}_\theta
\end{equation}
and
\begin{equation}\label{definition alpha lambdas}
    \gamma=\frac{3M^6(1+\beta)}{m^2}\,, \quad
    \gamma_1=3M^6\left(\frac{\rho+\rho_{\cal C}}{m^2}-\Lambda_{\rm RS}\right)\,,
    \quad \gamma_2=3M^6\left(\frac{p+\frac{1}{3}\rho_{\cal C}}{m^2}+\Lambda_{\rm RS}\right)\,,
\end{equation}
we obtain from \eqref{effective rewritten}:
\begin{eqnarray}
\label{1m}
    x^2-y^2-z^2+2y z = \gamma x+\gamma_1\,, \\
\label{2m}
     x^2-y^2+z^2+2x z = \gamma y+\gamma_2\,, \\
\label{3m}
    x^2+y^2+xy+xz-y z = \gamma z+\gamma_2\,,
\end{eqnarray}
which is a system of algebraic equations with the components of
$E^\alpha{}_\beta$ playing the role of unknown variables.

Because of the condition $C^r{}_r=C^\theta{}_\theta$, the same constant
$\gamma_2$ appears on the right-hand side of \eqref{2m} and \eqref{3m}. This
allows us to solve this system of algebraic equations in the most general case.
Subtracting \eqref{2m} from \eqref{3m}, we obtain
\begin{equation}\label{alternative}
    (y-z)(x+2y+z+\gamma)=0\,,
\end{equation}
which admits two alternative solutions: $y=z$ or $x+2y+z+\gamma=0$.

First, consider the case $y=z$. The solution of \eqref{1m}, \eqref{2m} is then:
\begin{equation}\label{einstein with matter y=z 00}
    - G^t{}_t-\frac{6(1+\beta)}{\ell^2}=\frac{1}{m^2}\rho \pm
    \frac{2}{\ell}\sqrt{3\left[\frac{\rho+\rho_{\cal C}}{m^2}+\frac{3(1+\beta)^2}{\ell^2}-\Lambda_{\rm RS}\right]}\,,
\end{equation}
\begin{equation}\label{einstein with matter y=z 11}
    G^r{}_r+\frac{6(1+\beta)}{\ell^2}=\frac{1}{m^2}p \mp
    \frac{6}{\ell}\cdot\frac{\left[\frac{\rho-p+2\rho_{\cal C}/3}{2m^2}+\frac{3(1+\beta)^2}{\ell^2}-\Lambda_{\rm RS}\right]}{\sqrt{3\left[\frac{\rho+\rho_{\cal C}}{m^2}+\frac{3(1+\beta)^2}{\ell^2}-\Lambda_{\rm RS}\right]}}\,,
\end{equation}
\begin{equation}\label{einstein with matter y=z 22}
     G^\theta{}_\theta +\frac{6(1+\beta)}{\ell^2}=\frac{1}{m^2}p \mp
    \frac{6}{\ell}\cdot\frac{\left[\frac{\rho-p+2\rho_{\cal C}/3}{2m^2}+\frac{3(1+\beta)^2}{\ell^2}-\Lambda_{\rm RS}\right]}{\sqrt{3\left[\frac{\rho+\rho_{\cal C}}{m^2}+\frac{3(1+\beta)^2}{\ell^2}-\Lambda_{\rm RS}\right]}}\,.
\end{equation}

The second solution of \eqref{alternative}, namely, $x+2y+z+\gamma=0$, leads to
the result
\begin{equation}\label{einstein with matter another 00}
   - G^t{}_t-\frac{6(1+\beta)}{\ell^2}=\frac{1}{m^2}\rho \mp
    \frac{2}{\ell}\cdot\frac{\left[\frac{\rho+3p+2\rho_{\cal C}}{2m^2}-\frac{3(1+\beta)^2}{\ell^2}+\Lambda_{\rm RS}\right]}{\sqrt{\frac{p+\rho_{\cal C}/3}{m^2}-\frac{3(1+\beta)^2}{\ell^2}+\Lambda_{\rm RS}}}\,,
\end{equation}
\begin{equation}\label{einstein with matter another 11}
    G^r{}_r+\frac{6(1+\beta)}{\ell^2}=\frac{1}{m^2}p \pm
    \frac{2}{\ell}\sqrt{\frac{p+\rho_{\cal C}/3}{m^2}-\frac{3(1+\beta)^2}{\ell^2}+\Lambda_{\rm RS}}\,,
\end{equation}
\begin{equation}\label{einstein with matter another 22}
   G^\theta{}_\theta +\frac{6(1+\beta)}{\ell^2}=\frac{1}{m^2}p \pm
    \frac{2}{\ell}\cdot\frac{\left[\frac{\rho-p+2\rho_{\cal C}/3}{2m^2}+\frac{3(1+\beta)^2}{\ell^2}-\Lambda_{\rm RS}\right]}{\sqrt{\frac{p+\rho_{\cal C}/3}{m^2}-\frac{3(1+\beta)^2}{\ell^2}+\Lambda_{\rm RS}}}\,.
\end{equation}

It is remarkable that the effective braneworld equations in a static
spherically symmetric situation in our model have the form of Einstein
equations with a special modification on the right-hand side. We should also
note that the system of equations \eqref{einstein with matter y=z
00}--\eqref{einstein with matter y=z 22} matches with the two branches of
cosmological equations \eqref{mimic} of the braneworld model derived in
\cite{cosmology}. This means that writing the components $G^\alpha{}_\beta$ for
the Friedmann--Robertson--Walker metric and substituting them to the left-hand
side of \eqref{einstein with matter y=z 00}--\eqref{einstein with matter y=z
22}, one obtains the cosmological equations \eqref{mimic}.  As for the branches
described by equations \eqref{einstein with matter another 00}--\eqref{einstein
with matter another 22}, their connection with cosmology is not clear.  Because
of this, we restrict our investigation only to the analysis of system
\eqref{einstein with matter y=z 00}--\eqref{einstein with matter y=z 22}.
Following the cosmological nomenclature, we call the branch with the lower
(`$-$') sign Brane\,1, and the branch with the upper (`$+$') sign Brane\,2.
These two branches have a geometric origin: they correspond to the two possible
branches of solutions in the five-dimensional bulk space \cite{cosmology}.

It is convenient to introduce new functions
\begin{equation}\label{varrho}
    \varrho=\rho\pm
    2\xi\rho_L\left(\sqrt{1+\frac{\rho+\rho_{\cal C}}{\rho_L}}-1\right)\,,
\end{equation}
\begin{equation}\label{P}
    {\cal P}=p\mp 2\xi\rho_L\left(\frac{1+\frac{\rho-p+2\rho_{\cal C}/3}{2\rho_L}}{\sqrt{1+\frac{\rho+\rho_{\cal C}}{\rho_L}}}-1\right)\,,
\end{equation}
and a new constant
\begin{equation}\label{lambda upper}
    \lambda=\frac{6}{\ell^2}\left(1+\beta\pm\frac{1}{\xi}\right)\,,
\end{equation}
where $\rho_L$ and $\xi$ are constant parameters defined by
\begin{equation}\label{rhoL xi}
    \rho_L=\frac{3m^2}{L^2}\,, \qquad \xi=\frac{L}{\ell}\,,
\end{equation}
and $L$ is a new length scale
\begin{equation}\label{L}
    \frac{1}{L^2}=\frac{1}{\ell^2}+\frac{\sigma}{3m^2}-\frac{\Lambda}{6}\,.
\end{equation}

We assume the expression on the right-hand side of \eqref{L} to be positive to
avoid singularity in the square root of \eqref{varrho}, \eqref{P}, hence, also
in system \eqref{einstein with matter y=z 00}--\eqref{einstein with matter y=z
22}, for positive values of $\rho$ and $\rho_{\cal C}$.\footnote{These
singularities directly correspond to the `quiescent' singularities in
braneworld cosmology discussed in \cite{TTSS}.} To avoid a similar singularity
in \eqref{einstein with matter another 00}--\eqref{einstein with matter another
22}, one would require the opposite condition, namely
$1/\ell^2+\sigma/3m^2-\Lambda/6<0$. Thus, one can think that the choice between
the two systems of equations \eqref{einstein with matter y=z
00}--\eqref{einstein with matter y=z 22} and \eqref{einstein with matter
another 00}--\eqref{einstein with matter another 22} is determined by the value
of the expression on the right-hand side of \eqref{L}. If this expression is
equal to zero, then these two systems coincide.

The signs of $\sigma$ (or $\beta$) and $\Lambda$ are not yet specified, but one
should note that the above condition leads to the constraint $\beta
> - 1/2 + \Lambda \ell^2/12$. Transition to the DGP model in our equations is
realized by setting $\sigma = 0$ and $\Lambda=0$, which implies $\beta=0$ and
$\xi=1$. In the cosmic-mimicry model with $\beta \approx - \alpha \approx -
5/4$, which we discussed in Sec.~\ref{sec:mimic}, the value of $\xi$ can be
found to be
\begin{equation}\label{xi in mimicry}
    \xi\approx\frac{1}{|1 + \beta|}\approx 4\,.
\end{equation}

In principle, the density of the Weyl fluid $\rho_{\cal C}$, unlike that of
realistic matter density, is not restricted in its sign.  However, the above
condition of absence of singularity restricts the range of boundary conditions
for $\rho_{\cal C}(r)$ which we shall specify in the next section.

In the new notation \eqref{varrho}--\eqref{lambda upper}, system
\eqref{einstein with matter y=z 00}--\eqref{einstein with matter y=z 22} is
written as follows:
\begin{equation}\label{G00 interior}
    - G^t{}_t=\lambda + \frac{1}{m^2}\varrho\,,
\end{equation}
\begin{equation}\label{G11 interior}
    G^r{}_r=-\lambda + \frac{1}{m^2}{\cal P}\,,
\end{equation}
\begin{equation}\label{G22 interior}
 G^\theta{}_\theta = -\lambda + \frac{1}{m^2}{\cal P}\,.
\end{equation}
The quantities $\varrho$ and ${\cal P}$ play the role of effective energy
density and pressure in the usual Einstein equations describing the static
spherically symmetric situation, and $\lambda$ is the effective cosmological
constant. The formal solution of \eqref{G00 interior}--\eqref{G22 interior} is
well known:
 \begin{equation}\label{interior metric}
    ds^2=-e^{2\phi(r)}dt^2+\frac{dr^2}{1-2\mu(r)/r}+r^2\left(d\theta^2+\sin^2\theta
    d\varphi^2\right)\,,
\end{equation}
where
\begin{equation}\label{mu interior}
    \mu(r)=\frac{\lambda
    r^3}{6}+\frac{1}{2m^2}\int^{r}_{0}\varrho(r')r'^2dr'\,,
\end{equation}
\begin{equation}\label{phi interior}
    \phi'(r)=\frac{\mu(r)+r^3{\cal P}(r)/2m^2 - \lambda
    r^3/2}{r\left[r-2\mu(r)\right]}\,.
\end{equation}

The distributions of the densities of usual matter and Weyl fluid are
determined by the energy--momentum conservation and by the analogue of the
Tolman--Oppenheimer--Volkoff equation [which, in our case, is a simple
consequence of the Bianchi identity applied to \eqref{G00 interior}--\eqref{G22
interior}]. In Einstein's theory, the energy--momentum conservation equation
and the Bianchi identity give the same result, which is precisely the
Tolman--Oppenheimer--Volkoff equation. In our model, due to the presence of
{\em effective\/} $\varrho$ and ${\cal P}$ on the right-hand side of \eqref{G00
interior}--\eqref{G22 interior}, which do not coincide with the usual $\rho$
and $p$, we have two different equations:
\begin{equation}\label{conservation equation interior}
    p'+\phi'(\rho+p)=0\,,
\end{equation}
\begin{equation}\label{TOV equation}
    \left(\frac{1}{3}+\zeta\right)\rho'_{\cal C}+(1+\zeta)\rho'=\frac{8\rho_{\cal C}}{3}\phi'\,,
\end{equation}
where
\begin{equation}\label{gamma}
    \zeta=\frac{p+\rho_{\cal C}/3-\rho_L}{\rho+\rho_{\cal C}+\rho_L}\,.
\end{equation}

For a given equation of state $p=p(\rho)$ and some boundary conditions for
$\rho(r)$ and $\rho_{\cal C}(r)$, one can integrate (at least, numerically)
\eqref{mu interior}, \eqref{phi interior}, \eqref{conservation equation
interior}, and \eqref{TOV equation} to obtain the metric and matter density
distribution inside a massive object.  As for the boundary conditions, it looks
natural to specify the values of $\rho$ and $\rho_{\cal C}$ at the center of
the spherically symmetric gravitating body. At the surface of this object, the
interior solution matches with the exterior vacuum solution of the above
equations.

It should be noted here that Eq.~\eqref{TOV equation} gives the
condition of incompressible fluid $\rho'(r)=0$ for the ordinary
matter if we set $\rho_{\cal C}(r)=0$, which corresponds to a
complete neglect of the tensor $C_{ab}$ in the initial field
equations \cite{SMS,KPP}. Allowing for nonzero density of the Weyl
fluid replaces this restrictive condition on $\rho(r)$ by a
differential equation determining the behavior of $\rho_{\cal
C}(r)$.

We see from \eqref{G00 interior}--\eqref{G22 interior} that the
effective gravitational constant is given by $8\pi G_{\rm
N}=1/m^2$. For $\ell \approx 2$~Mpc and $\xi\approx 4$, the value
$\rho_L$ in \eqref{rhoL xi} can be estimated to be $\rho_L\approx
10^{-22}~\mbox{g}\cdot \mbox{cm}^{-3}$. This density scale lies
far below the density of compact objects, such as stars and
planets. Thus, inside such objects, we expect only slight
modification of Newtonian dynamics due to the bulk effects
[$\varrho(r)\approx\rho(r)$ and ${\cal P}(r)\approx p(r)$ if
$\rho_{\cal C}(r)\lesssim 1$]. The vacuum situation corresponding
to this case will be studied in the next section.

The local disk density of a typical spiral galaxy is $\rho_{\rm disk}\sim
3$$-$$12\times10^{-24}~\mbox{g}\cdot \mbox{cm}^{-3}$, and, therefore, the bulk
effects inside the disk are significant. The solution of the interior problem
for such objects is of special interest, but is beyond the target of this
article. The solution to this problem will hopefully give a relation between
the asymptotic rotational velocity $\upsilon_c$ and the gravitational mass of a
spiral galaxy.


\section{Vacuum solutions}
\label{sec:vacuum}

\subsection{General equations and boundary condition}

In the vacuum, which is defined by the conditions $\rho(r)=0$ and $p(r)=0$, the
effective matter density $\varrho(r)$ and pressure ${\cal P}(r)$ are not
necessarily equal to zero due to the possible presence of the Weyl fluid
component $\rho_{\cal C}(r)\,$:
\begin{equation}\label{varrho vac}
    \varrho_{\rm vac}(r)=\pm
    2\xi\rho_L\left(\sqrt{1+\rho_{\cal C}(r)/\rho_L}-1\right)\,,
\end{equation}
\begin{equation}\label{P vac}
    {\cal P}_{\rm vac}(r)=\mp
    2\xi\rho_L\left(\frac{1+\rho_{\cal C}(r)/3\rho_L}{\sqrt{1+\rho_{\cal C}(r)/\rho_L}}-1\right)\,.
\end{equation}

One should note that the effective equation of state in this case
\begin{equation}\label{effective equation of state}
    \frac{{\cal P}_{\rm vac}}{\varrho_{\rm vac}}=\frac{1}{3}\left(\frac{2}{\sqrt{1+\rho_{\cal C}/\rho_L}}-1\right)=\frac{1}{3}\cdot\frac{1\mp\varrho_{\rm vac}/(2\xi\rho_L)}{1\pm\varrho_{\rm vac}/(2\xi\rho_L)}
\end{equation}
differs from that of radiation.  Specifically, ${\cal P}_{\rm vac}/\varrho_{\rm
vac}<1/3$ if $\rho_{\cal C}(r)$ is positive.

In addition to equations \eqref{G00 interior}, \eqref{G11
interior}, we can use Eq.~\eqref{TOV equation} instead of
\eqref{G22 interior} in the vacuum case:
\begin{equation}\label{rho W differential y=z}
    \frac{\rho_{\cal C}-\rho_L}{\rho_{\cal C}+\rho_L}\,\rho_{\cal C}'=4\rho_{\cal C}\phi'\,.
\end{equation}

The key issue concerns the choice of the boundary condition for $\rho_{\cal
C}(r)$. As we noted above, it is natural to specify the value of $\rho_{\cal
C}$ at the center of the spherically symmetric gravitating body. However, as
the interior problem is rather complicated and because we are not going to
consider it in this paper, it is reasonable to parameterize the boundary
condition by the value of $\rho_{\cal C}$ in the vacuum at some radius $R$. For
a massive body, this could be the radius of its surface, from which the vacuum
solution starts. Thus $\rho_R\equiv\rho_{\cal C}(R)$ is a new free parameter of
our model, defining the boundary condition for the Eq.~\eqref{rho W
differential y=z} in vacuum. It is restricted by the condition
$\rho_R>-\rho_L$, which is required for the expression under the square root of
the right-hand side of \eqref{varrho vac} and \eqref{P vac} to be positive.

Eq.~\eqref{rho W differential y=z} can be solved exactly, relating
the energy density of the Weyl fluid $\rho_{\cal C}(r)$ and the
function $\phi(r)$ entering metric \eqref{interior metric}:
\begin{equation}\label{rho W y=z}
     \frac{\rho_{\cal C}(r)}{\left[\rho_L+\rho_{\cal C}(r)\right]^2}
     =\frac{\rho_R}{(\rho_L+\rho_R)^2}e^{4\left[\phi_R-\phi(r)\right]}\,,
\end{equation}
where $\phi_R\equiv\phi(R)$.  If the boundary value $\rho_R=0$,
then $\rho_{\cal C}(r) \equiv 0$ for $r \ge R$. In this case,
$\varrho_{\rm vac}(r) = {\cal P}_{\rm vac}(r)=0$ [see
\eqref{varrho vac} and \eqref{P vac}], and we obtain the usual
vacuum Einstein equations with a cosmological constant $\lambda$.
Thus, our braneworld model admits a Schwarzschild-(A)$dS_4$ space
as an exact vacuum solution. The interior counterpart of this
exterior solution in the context of the Randall--Sundrum
braneworld model was analyzed in \cite{GM,O}.

To solve equations \eqref{G00 interior} and \eqref{G11 interior} in the vacuum,
we introduce the function
\begin{equation}\label{m and delta}
    \Delta(r)=\frac{e^{2\phi(r)}}{1-2\mu(r)/r}\,.
\end{equation}
Then one has
\begin{equation}\label{vacuum einstein y=z m}
     \frac{2\mu'}{r^2}=\lambda\pm
    \frac{2\xi\rho_L}{m^2}\left(\sqrt{1+\frac{\rho_{\cal C}}{\rho_L}}-1\right)\,,
\end{equation}
\begin{equation}\label{vacuum einstein y=z delta}
     \frac{1}{r}\left(1-\frac{2\mu}{r}\right)
     \frac{\Delta'}{\Delta}=\pm\frac{4\xi}{3m^2}\frac{\rho_{\cal C}}{\sqrt{1+\rho_{\cal C}/\rho_L}}\,.
\end{equation}
This nonlinear system in its full generality looks quite
complicated. But, in fact, one can note that the gravitational
field is usually weak in the neighborhood of an astrophysical
object. So, in analysis of \eqref{vacuum einstein y=z m},
\eqref{vacuum einstein y=z delta}, we restrict ourselves only by
the Newtonian approximation.

\subsection{Newtonian approximation}

First, we note that the function $\phi(r)$ is an analogue of the
Newtonian gravitational potential in this case. Under the
condition
\begin{equation}\label{newtonian limit}
    |\phi(r)|\ll 1 \, ,
\end{equation}
Eq.~\eqref{rho W y=z} implies an almost constant value of
$\rho_{\cal C}(r)\,$:
\begin{equation}\label{rho W Newton}
    \rho_{\cal C}(r)\approx\rho_R\,.
\end{equation}
In this approximation, Eq.~\eqref{vacuum einstein y=z m} can be
integrated:
\begin{equation}\label{newt limit mu}
   1-\frac{2\mu(r)}{r}=1-\frac{r_g}{r}-\frac{\lambda r^2}{3}\mp\frac{2\xi r^2}{L^2}\left(\sqrt{1+\frac{\rho_R}{\rho_L}}-1\right)\,,
\end{equation}
where $r_g$ is the integration constant and is an analogue of the
Schwarzschild radius in general relativity.

The Newtonian approximation requires also the condition $|\mu(r)/r |\ll 1$.
Assuming it to be satisfied, we can solve Eq.~\eqref{vacuum einstein y=z
delta}:
\begin{equation}\label{newt limit delta}
    \Delta(r)\approx e^{\pm r^2/r_{\rm N}^2}\approx
    1\pm\frac{r^2}{r_{\rm N}^2}\,,
\end{equation}
where
\begin{equation}\label{newt r0}
   r_{\rm N}^2=\frac{L^2}{2\xi}\cdot\frac{\sqrt{1+\rho_R/\rho_L}}{|\rho_R/\rho_L|}\,
\end{equation}
determines the radius at which the Newtonian approximation fails. The
integration constant that arises in solving \eqref{vacuum einstein y=z delta}
is absorbed by rescaling the time coordinate. The parameter $r_{\rm N}$ is
unambiguously defined by $\rho_R$ and can be taken as a new free parameter of
our model. The Newtonian potential in this case is defined by
\begin{equation}\label{newt limit potential}
    e^{2\phi(r)}\approx1+2\phi(r)\approx
    1-\frac{r_g}{r}-\frac{\lambda r^2}{3}\pm\frac{2\xi r^2}{L^2}\cdot\frac{\left(\sqrt{1
    +\rho_R/\rho_L}-1\right)}{\sqrt{1+\rho_R/\rho_L}}\,,
\end{equation}
and the region of applicability of the above result is
\begin{equation}\label{newton applicability}
    r_g\ll r\ll r_{\rm N}\,.
\end{equation}

One can easily verify that, under condition \eqref{newton applicability}, our
initial assumptions $|\phi(r)|\ll 1$ and $|\mu(r)/r| \ll 1$ are satisfied. If
$\rho_R/\rho_L$ is of order unity, then $r_{\rm N}\sim L/\sqrt{\xi}$. Equations
\eqref{newt limit mu} and \eqref{newt limit potential} then describe the
Schwarzschild metric in the region $r^3\lesssim \xi r_g\ell^2$. For $\xi$ given
by \eqref{xi in mimicry}, this condition is compatible with condition
\eqref{scale}, thus justifying our initial assumption that bulk effects is
insignificant at sufficiently small scales.

A remarkable feature of our model in comparison with the
Randall--Sundrum braneworld model is the appearance of a new
length scale $\ell$ [which goes to zero in the limit $m \to 0$;
see \eqref{L}] defining the radius $r_*\sim r_g \ell^2$ [see
\eqref{scale}] up to which the Newtonian gravity works. Beyond the
radius $r_*$, our model predicts transition to gravity with
different properties, for which the influence of the projection of
the bulk Weyl tensor $C_{ab}$ on the dynamics can be significant,
thus allowing to explain the flatness of galactic rotation curves,
as we have demonstrated in Sec.~\ref{sec:rec}.


\section{Conclusion}
\label{sec:conclusion}

In this paper, we continued the analysis of a generic braneworld model with
induced gravity described by action \eqref{action}.  It was shown previously in
\cite{SSV} that a low-density braneworld can mimic the expansion properties of
the LCDM model. In particular, a universe consisting solely of baryons with
$\Omega_{\rm b} \simeq 0.04$ can mimic the LCDM cosmology with a much larger
`effective' value of the matter density $\Omega_{\rm m}^{\rm LCDM} \simeq
0.2$$-$$0.3$.  This property fixes the parameter $\beta$ defined in
\eqref{beta} to be $\beta \approx - 5/4$.

The general analysis of the braneworld model with induced gravity performed in
Sec.~\ref{sec:properties} demonstrates that, in the neighborhood of the source,
at distances smaller than $r_*$ given by \eqref{scale}, the solution is
determined mainly by the quadratic part $Q_{ab}$ in \eqref{effective}, which
means that it respects the `bare' Einstein equation $m^2 G_{ab} - T_{ab}=0$ to
a high precision. This effect is sometimes described as the `gravity filter' of
the DGP model \cite{DGP}, which screens the scalar graviton in the neighborhood
of the source making the gravity effectively Einsteinian.

At large distances, the modified gravitational field equations can be used to
explain the properties of the galactic rotation curves. Using the smallness of
the rotational velocity $\upsilon_c$, we have developed a perturbative scheme
for reconstructing the metric in this region. In the leading order of
expansion, in the region given by \eqref{condition of pert}, our result
reproduces that obtained for the Randall--Sundrum braneworld model in
\cite{HC,BH}. The vacuum gravity in this case can be approximately described by
the Einstein equations with the traceless projection $C_{ab}$ of the bulk Weyl
tensor to the brane playing the role of some effective stress--energy tensor of
galactic fluid \eqref{halo fluid}. The left inequality in \eqref{condition of
pert} is satisfied in a real spiral galaxy such as our Milky Way with the
distance $r \sim 3$~kpc at which the rotation velocity becomes constant,
$\upsilon_c \sim 7\times 10^{-4}$, if $\ell \lesssim \sqrt{|1 + \beta|}\,
\upsilon_c^{-1}\, r \simeq 2$~Mpc.  In this paper, we restricted ourselves to
this case only.  In the opposite case, one has to solve a nonlinear
differential equation \eqref{g}.

The role of dark matter in the braneworld model is played by the traceless
tensor of the Weyl fluid. Using the methods of \cite{FV}, in which it was
proposed to use the existing data on rotation curves and lensing measurements
to constrain the equation of state of galactic fluid, we have shown that
deflection angles in our model are about $75\%$ of the usual predictions based
on dark matter. This result confirms the analysis done in \cite{PBK} and allows
for a direct verification of the braneworld paradigm.

In the region near to the gravitational object, we derive a closed system of
equations for a static spherically symmetric situation in the case of zero
anisotropic stress in the tensor $C_{ab}$. We demonstrate that the effective
braneworld equations in this case have the form of Einstein equations with a
special modification on the right-hand side [see \eqref{G00
interior}--\eqref{G22 interior}]. We find the Schwarzschild metric to be an
approximate vacuum solution of these equations at distances well below the
scale $r_*$, in agreement with the general consideration of
Sec.~\ref{sec:properties}.

Expression \eqref{scale} for $r_*$ generalizes the length scale \cite{vdvz} of
the DGP model, below which nonlinear effects become important, to the case of
nonzero brane tension (nonzero $\beta$) and bulk cosmological constant
satisfying the Randall--Sundrum constraint $\Lambda_{\rm RS}=0$. For the
parameter values $\beta \approx - 5/4$ and $\ell \lesssim 2$~Mpc determined
from the explanation of dark-matter phenomena, the value of $r_*$ for the Sun
turns out to be of the order of $1$~pc, which is quite large so that the
solar-system experiments are in accord with the general-relativistic
expectations.

The density scale \eqref{rhoL xi} which appears in our model is estimated to be
$\rho_L\approx 10^{-22}~\mbox{g}\cdot \mbox{cm}^{-3}$. This value lies far
below the density of compact objects, such as stars and planets. Thus, inside
such objects, we expect only slight modification of Newtonian dynamics due to
the bulk effects. However, the local disk density of a typical spiral galaxy is
$\rho_{\rm disk}\sim 3$$-$$12\times10^{-24}~\mbox{g}\cdot \mbox{cm}^{-3}$, and,
therefore, the bulk effects inside the disk should be significant. The interior
problem for such objects remains a challenge for future investigation. The
solution to this problem will hopefully give a relation between the asymptotic
rotational velocity $\upsilon_c$ and the mass of a spiral galaxy. It may help
to verify the Tully--Fisher relation and extend our results to a wider class of
astrophysical objects such as ellipticals and ultra compact dwarf galaxies.


\section*{Acknowledgments}

The authors acknowledge valuable discussions with Naresh Dadhich and Varun
Sahni.  This work was supported in part by the INTAS grant No.~05-1000008-7865.


\end{document}